%% file: RR-6446.tex
\documentclass[a4paper]{article}

\usepackage{algorithm}
\usepackage{graphicx}
\usepackage{amssymb}
\usepackage[isolatin]{inputenc}
\usepackage[OT1]{fontenc}   
\usepackage{dsfont}
\usepackage{epsfig}
\usepackage{RR}
\RRNo{6446}

\input{head}

\RRNo{9999}

\RRdate{February 2008}

\RRtitle{Stabilisation instantanée dans les systèmes à passage de messages}
\RRetitle{Snap-Stabilization in Message-Passing Systems}

\titlehead{Snap-Stabilization in Message-Passing Systems}

\RRauthor{Sylvie Delaët\thanks{Université Paris-Sud, France}
\and
Stéphane Devismes\thanks{CNRS, Université Paris-Sud, France}
\and
Mikhail Nesterenko\thanks{Computer Science Department, Kent State University, USA}
\and
Sébastien Tixeuil\thanks{Université Paris 6, LIP6-CNRS \& INRIA, France}
}

\RRresume{Dans cet article, nous considérons le problème, jusqu'ici ouvert, de la stabilisation instantanée dans les systèmes à passage de messages. La stabilisation instantanée est une approche élégante permettant de réaliser des protocoles qui supportent les fautes transitoires. Par rapport à l'approche auto-stabilisante, la stabilisation instantanément stabilisante assure que l'effet des fautes est contenu immédiatement après que celles-ci cessent. Notre contribution est double: nous prouvons que (1) la stabilisation instantanée est impossible pour de nombreux problèmes si nous supposons des réseaux où la capacité des canaux de communications est finie mais non bornée; (2) la stabilisation instantanée devient possible avec les mêmes paramètres si on suppose que la capacité des canaux est bornée. A titre d'exemple, Nous proposons trois protocoles instantanément stabilisants fonctionnant dans un réseau complet. Ces travaux ouvrent de nouvelles perspectives de recherche car ils démontrent que la stabilisation instantanée peut être implantée dans les réseaux actuels.} 

\RRabstract{In this paper, we tackle the open problem of snap-stabilization in message-passing systems. Snap-stabilization is a nice approach to design protocols that  withstand transient faults. Compared to the well-known self-stabilizing approach, snap-stabilization guarantees that the effect of faults is contained immediately after faults cease to occur. Our contribution is twofold: we show that (1) snap-stabilization is impossible for a wide class of problems if we consider networks with finite yet unbounded channel capacity; (2) snap-stabilization becomes possible in the same setting if we assume bounded-capacity channels. We propose three snap-stabilizing protocols working in fully-connected networks. Our work opens exciting new research perspectives, as it enables the snap-stabilizing paradigm to be implemented in actual networks.}

\RRmotcle{Syst\`{e}mes distribu\'{e}s, Algorithme distribu\'{e}, Auto-stabilisation, Stabilisation Instantanée}
\RRkeyword{Distributed systems, Distributed algorithm, Self-stabilization, Snap-Stabili\-za\-tion}

\RRprojet{Grand large}
\RRtheme{\THNum}
\URFuturs

\begin{document}

\makeRR

\input{intro}

\input{model}

\input{def}

\input{spec}

\input{infty}

\section{Snap-Stabilizing Message-Passing Protocols}\label{sect:algo}

We now consider systems with channels having a bounded capacity. In such systems, we assume that if a process sends a message in a channel that is full, then the message is lost. We restrict our study to systems with single-message capacity channels. The extention to an arbitrary but known bounded message capacity is straightforward (see \cite{APV91,AKMPV07}). We propose three snap-stabilizing protocols (Algorithms \ref{algo:PIF}-\ref{algo:ME}) for the {\em Propagation of Information with Feedback} (PIF), IDs-Learning, and mutual exclusion problem, respectively. The PIF is a basic tool allowing us to solve the two other problems. The IDs-Learning is a simple application of the PIF. Finally, the mutual exclusion protocol uses the two former protocols. 

\input{pif}

\input{id}

\input{me}

\input{ccl}

\bibliographystyle{plain}
\bibliography{bibliothese}

\end{document}

%% file: head.tex
\newtheorem{definition}{Definition}
\newtheorem{theorem}{Theorem}
\newtheorem{lemma}{Lemma}

\newtheorem{hypothesis}{Hypothesis}

\newtheorem{corollary}{Corollary}
\newtheorem{specification}{Specification}
\newtheorem{property}{Property}

\newenvironment{proof}{{\bf Proof. } }{{\hfill $\Box$}\vspace{.5pc}}

\newcommand{\led}{\mathcal{L}}
\newcommand{\req}{\mathtt{Request}}
\newcommand{\rwait}{\mathtt{Wait}}
\newcommand{\rin}{\mathtt{In}}
\newcommand{\rdone}{\mathtt{Done}}
\newcommand{\state}{\mathtt{State}}
\newcommand{\neigstate}{\mathtt{NeigState}}

\newcommand{\IDL}{\mathcal{IDL}}
\newcommand{\rzero}{\mathtt{A}_0}
\newcommand{\rone}{\mathtt{A}_1}
\newcommand{\rtwo}{\mathtt{A}_2}
\newcommand{\rthree}{\mathtt{A}_3}
\newcommand{\rfour}{\mathtt{A}_4}
\newcommand{\rfive}{\mathtt{A}_5}
\newcommand{\rsix}{\mathtt{A}_6}
\newcommand{\rseven}{\mathtt{A}_7}
\newcommand{\reight}{\mathtt{A}_8}
\newcommand{\rnine}{\mathtt{A}_9}
\newcommand{\rten}{\mathtt{A}_{10}}
\newcommand{\C}{\mathcal{C}}
\newcommand{\Syst}{\mathcal{S}}
\newcommand{\I}{\mathcal{I}}
\newcommand{\Min}{\mathtt{minID}}
\newcommand{\IDtab}{\mathtt{ID}\mbox{-}\mathtt{Tab}}
\newcommand{\midc}{\mathtt{IDL}}
\newcommand{\pif}{\mathcal{PIF}}
\newcommand{\me}{\mathcal{ME}}
\newcommand{\mes}{\mathtt{B}\mbox{-}\mathtt{Mes}}
\newcommand{\fmes}{\mathtt{F}\mbox{-}\mathtt{Mes}}
\newcommand{\pifm}{\mathtt{PIF}}
\newcommand{\ok}{\mathtt{OK}}
\newcommand{\phase}{\mathtt{Phase}}
\newcommand{\val}{\mathtt{Value}}
\newcommand{\nval}{\mathtt{Privileges}}
\newcommand{\ask}{\mathtt{ASK}}
\newcommand{\exit}{\mathtt{EXIT}}
\newcommand{\exitcs}{\mathtt{EXITCS}}
\newcommand{\yes}{\mathtt{YES}}
\newcommand{\no}{\mathtt{NO}}
\newcommand{\IF}{{\bf if\ }}
\newcommand{\THEN}{{\bf then}}
\newcommand{\ELSE}{{\bf else}}
\newcommand{\FI}{{\bf end if}}
\newcommand{\FOR}{{\bf for all\ }}
\newcommand{\DO}{{\bf do}}
\newcommand{\DONE}{{\bf done}}
\newcommand{\SEND}{{\bf send}}
\newcommand{\RECV}{{\bf receive}}
\newcommand{\BRD}{{\bf receive\mbox{-}brd}}
\newcommand{\FCK}{{\bf receive\mbox{-}fck}}
\newcommand{\FROM}{{\bf from\ }}
\newcommand{\TO}{{\bf to\ }}
\newcommand{\cs}{\mathtt{\langle CS \rangle}}
\newcommand{\Rem}[1]{\qquad $\slash *$ #1 $* \slash$}
\newcommand{\N}{\mathds{N}}
\newcommand{\IFF}{{\em if and only if\ }}
\newcommand{\wrt}{{\em w.r.t.\ }}

\newcommand{\BEGLIST}{\begin{list}{}{\partopsep -2pt \parsep -2pt \listparindent 0pt \labelwidth .5in}}
\newcommand{\ENDLIST}{\end{list}}

%% file: intro.tex
\section{Introduction}\label{sect:intro}

\emph{Self-stabilization} \cite{Dij74} is an elegant approach to
forward failure recovery. 
Regardless of the global state to which the failure drives the system,
after the influence of the failure stops, a self-stabilizing system is
guaranteed to resume correct operation.  This guarantee comes at the
expense of temporary safety violation. That is, a self-stabilizing
system may behave incorrectly as it recovers. Bui~\emph{et
al}~\cite{BDPV99b} introduce a related concept of
\emph{snap-stabilization}. Given a problem specification, a system is
guaranteed to perform according to this specification regardless of
the initial state. If the system is sensitive to safety violation
snap-stabilization becomes an attractive option.
However, the snap-stabilizing protocols presented thus far assume a
rather abstract shared memory model. In this model a process reads the
states of all of its neighbors and updates its own state in a single
atomic step. The protocol design with forward recovery mechanisms such
as self- and snap-stabilization under more concrete program model such
as asynchronous message-passing is rather challenging. As Gouda and
Multari~\cite{DBLP:journals/tc/GoudaM91} demonstrate,  
if channels can hold an arbitrary number of messages, a large number
of problems could not be solved by self-stabilizing algorithms: a
pathological corrupted state with incorrect messages in the channels may
prevent the protocol from stabilizing. See also Katz and
Perry~\cite{KP93a} for additional detail on this topic. The issue is
exacerbated for snap-stabilization by the stricter safety
requirements.  Thus, however attractive the concept, the applicability
of snap-stabilization to concrete models, such as message-passing
models remained.  In this paper we address this problem. We outline
the bounds of the achievable and present snap-stabilizing solutions in
message-passing systems for several practical problems.

\paragraph{Related literature.} Several studies modify the
concept of self-stabilization to add safety property during recovery
from faults. Dolev and Herman~\cite{dolev97superstabilizing} introduce
\emph{super-stabilization} where a self-stabilizing protocol can recover
from a local fault while satisfying a safety predicate. This theme is
further developed as \emph{fault-containment}~\cite{GGHP00}.

A number of snap-stabilizing protocols are presented in the
literature. In particular \emph{propagation of information with
feedback}(PIF) is a popular problem to
address~\cite{BDPV99b,BDPV99c,DBLP:journals/dc/BuiDPV07,DBLP:journals/jhsn/CournierDPV05,CDPV02,BCV03,CDV06}. Several
studies present snap-stabilizing token circulation
protocols~\cite{DBLP:journals/jpdc/PetitV07,CDPV2006,CDV05}. There
also exists snap-stabilizing protocols for neighborhood
synchronization~\cite{JADT02j}, binary search tree construction
\cite{DBLP:conf/sss/BeinDV05} and cut-set detection \cite{CDV205}.
Cournier \emph{et al}~\cite{CDPV2003} propose a method
to add snap-stabilization to a large class of protocols. 

Unlike snap-stabilization, self-stabilizing protocol were designed for
message-passing systems of unbounded capacity channels.  Afek and
Brown~\cite{AB93} use a string of random sequence numbers to
counteract the problem of infinite-capacity channels and design a
self-stabilizing \emph{alternating-bit} protocol (ABP). Dela{\"e}t
\emph{et al}~\cite{DDT06c} propose a method to design self-stabilizing
protocols for a class of terminating problems in message-passing
systems with lossy channels of unbounded capacity. Awerbuch \emph{et
al}~\cite{APV91} describe the property of \emph{local correctability}
and demonstrate who to design locally-correctable self-stabilizing
protocols. Researchers also consider message-passing systems with
bounded capacity
channels~\cite{AB97,DBLP:journals/siamcomp/Varghese00,638828,DBLP:journals/dc/AroraN05,AKMPV07}.

\paragraph{Our contribution.}
In this paper, we address the problem of {\em snap-stabilization} in
message-passing systems. We introduce the concept of
\emph{safety-distributed problem specification} that encompasses most
practical problems and show that it is impossible to satisfy by a
snap-stabilizing protocol in message-passing systems with unbounded
finite channel capacity. That is if the channel capacity bound is
unknown to the processes. As a constructive contribution, we show that
snap-stabilization becomes possible if bound for the channel capacity
is known. We present the snap-stabilizing protocols that solve the
PIF, the
ID-learning and the mutual exclusion problems. To the best of our
knowledge these are the first snap-stabilizing protocols in such a
concrete program model.

\paragraph{Paper outline.} The rest of the paper is organized as follows.
We define the message-passing program model in
Section~\ref{sect:model}. In the same section, we describe the notion of snap-stabilization
and problem specifications.  In Section \ref{sect:infty}, we prove
the impossibility of snap-stabilization in message-passing systems
with channels of infinite capacity. We present the snap-stabilizing
algorithms for the system with bounded capacity channels in
Section~\ref{sect:algo}. We conclude the paper in
Section~\ref{sect:ccl}.

%% file: model.tex
\section{The Model}\label{sect:model}

We consider distributed systems having a {\em finite number of processes} and a {\em fully-connected topology}: any two distinct processes can communicate together by sending messages through a bidirectionnal link ({\em i.e.}, two channels in the opposite direction).

A process is a sequential deterministic machine that uses a local memory, a local algorithm, and input/output capabilities. Intuitively, such a process executes a local algorithm. This algorithm modifies the state of the process memory, and sends/receives messages through channels.

We assume that the channels incident to a process are locally distinguished by a {\em channel number}. For sake of simplicity, we assume that every process numbers its channels from 1 to $n-1$ ($n$ being the number of processes). In the following, we will indifferently use the notation $q$ to designate the process $q$ or the local channel number of $q$ in the code of some process $p$. We assume that the channels are FIFO but not necessary {\em reliable} (messages can be lost). However they all satisfy the following property: if an origin process $o$ sends infinitely many messages to a destination process $d$, then infinitely many messages are eventually received by $d$ from $o$. Also, we assume that any message that is never lost is received in a finite (but unbounded) time. 

The messages are of the following form: $\langle mes\-sage\mbox{-}ty\-pe$,$mes\-sage\mbox{-}va\-lue \rangle$. The $mes\-sage\mbox{-}va\-lue$ field is omitted if the message does not carry any value. The messages can contain more than one $message\mbox{-}value$.

An protocol consists of a collection of actions. An action is of the following form: $\langle label \rangle\ ::\ \langle guard \rangle\ \to\ \langle statement \rangle$. A {\em guard} is a boolean expression over the variables of a process and\slash or an {\em input} message. A {\em statement} is a sequence of assignments and\slash or message {\em sendings}. An action can be executed only if its guard is true. We assume that the actions are atomically executed, meaning that the evaluation of the guard and the execution of the corresponding statement of an action, if executed, are done in one atomic step. An action is said {\em enabled} when its guard is true. When several actions are simultaneously enabled at a process $p$, all these actions are sequentially executed following the order of their appearance in text of the protocol.

We reduce the {\em state} of each process to the state of its local memory, and the state of each link to its content. Hence, the global state of the system, referred to as {\em configuration}, can be simply defined as the product of the states of the memories of processes and of the contents of the links. 

A distributed system can be described using a {\em transition system} \cite{T01}. A {\em transition system} is a 3-uple $\Syst$ $=$ $(\C$, $\mapsto$,$\I)$ such that: $\C$ is set of configurations, $\mapsto$ is a binary transition relation on $\C$, and $\I \subseteq \C$ is the set of initial configurations. Using the notion of transition system, we can modelize the executions of a distributed system as follows: an \emph{execution} of $\Syst = (\C$,$\mapsto$,$\I)$ is a \emph{maximal} sequence of configurations $\gamma_0$, \dots, $\gamma_{i-1}$, $\gamma_i$, \dots\ such that: $\gamma_0 \in \I$ and $\forall i > 0$, $\gamma_{i-1} \mapsto \gamma_i$ ($\gamma_{i-1} \mapsto \gamma_i$ is referred to as a \emph{step}). In this paper, we only consider systems $\Syst = (\C$,$\mapsto$,$\I)$ such that $\I = \C$.

%% file: def.tex
\paragraph{Snap-Stabilization.}

\noindent In the following, a {\em specification} is a predicate defined on the executions.

\begin{definition}[Snap-Stabilization \cite{BDPV99b}]\label{def:snap} Let $\mathcal{SP_T}$ be a specification. An protocol $\mathcal P$ is {\em snap-sta\-bi\-li\-zing} for $\mathcal{SP_T}$ \IFF starting from any configuration, any execution of $\mathcal P$ satisfies $\mathcal{SP_T}$. 
\end{definition}

\noindent It is important to note that a snap-stabilizing protocol does not guarantee that the system never works in a fuzzy manner. Actually, the main idea behind the snap-stabilization is the following: the protocol is seen as a {\em function} and the function ensures two properties despites the arbitrary initial configuration of the system: (1) Upon an {\em external} (\wrt the protocol) {\em request} at a process $p$, the process $p$ (called the {\em initiator}) starts a {\em computation} of the function in finite time using special actions called {\em starting actions}. (2) If the process $p$ starts an {\em computation}, then the computation performs an {\em expected task}. With such properties, the protocol always satisfies its specifications. Indeed, when the protocol receives a request, this means that an external application (or a user) requests the computation of a specific task provided by the protocol. In this case, a snap-stabilizing protocol guarantees that the requested task is executed as expected. On the contrary, when there is no request, there is nothing to guarantee\footnote{This latter point is the basis of many misunderstandings about snap-stabilization. Indeed, due to the arbitrary initial configuration, some computations may initially run in the system without having been started: of course, snap-stabilization does not provide any guarantee on these non-requested computations. Consider, for instance, the problem of {\em mutual exclusion}. Starting from any configuration, a snap-stabilizing protocol cannot prevent several (non-requesting) processes to execute the critical section simultaneously. However, it guarantees that every requesting process executes the critical section in an exclusive manner.}.

%% file: spec.tex
\paragraph{Specifications.}

\noindent Due to the {\em Start} and {\em Correctness} properties it has to ensure, snap-sta\-bi\-li\-za\-tion requires specifications based on a sequence of actions (request, start, \dots) rather than a particular subset of configurations ({\em e.g.}, the {\em legitimate configurations}). Hence, for any task $\mathcal T$, we consider specifications of the following form:
\BEGLIST
\item[-] When requested, an {\em initiator} starts a computation of $\mathcal T$ in a finite time. {\bf (Start)}
\item[-] Any computation of $\mathcal T$ that is started is correctly performed. {\bf (Correctness)}
\ENDLIST

\smallskip

\noindent In this paper, the two first protocols we present are of a particular class: the {\em wave} protocols \cite{T01}. The particularity of such protocols is that they compute tasks that are {\em finite} and each of their computations contains at least one {\em decision event} that causally depends on an action at each process. Hence, our specifications for wave protocols contain two additionnal requirements:
\BEGLIST
\item[-] Each computation (even non-started) terminates in finite time. {\bf (Termination)}
\item[-] When the protocol terminates, if a computation was started, then at least one decision occurred and such a decision causally depends on an action at every process. {\bf (Decision})
\ENDLIST 

\paragraph{Self- {\em vs.} Snap-Stabilization. } Snap-stabilizing protocols are often compared to the self-stabilizing protocols --- such protocols converge in a finite time to a specified behavior starting from any initial configuration (\cite{Dij74}). The main advantage of the snap-stabilizing approach compared to the self-stabilizing one is the following: while a snap-stabilizing protocol ensures that any request is satisfied despite the arbitrary initial configuration, a self-stabilizing protocol often needs to be repeated an unbounded number of times before guarantying the proper processing of any request.

%% file: infty.tex
\section{Impossibility of Snap-Stabilization in Mes\-sa\-ge-Pas\-sing with Unbounded Ca\-pa\-ci\-ty Channels}\label{sect:infty}

\noindent In \cite{AlpernS85}, Alpern and Schneider observe that a specification is an intersection of {\em safety} and {\em liveness} properties. In \cite{AlpernS87}, the same authors define a {\em safety} property as a set of ``bad things'' that must never happen. Hence, it is sufficient to show that a prefix of an execution contains a ``bad thing'' to prove that the execution (and so the protocol) violates the safety property. We now consider {\em safety-distributed} specifications, {\em i.e.}, specifications having some {\em safety-distributed} properties. Roughly speaking, a {\em safety-distributed} property is a safety property that does not only depend on the behavior of a single process: some local behaviors at some processes are forbidden to be executed simultaneously while they are possible and do not violate the safety-distributed property if they are executed alone. For example, in the mutual exclusion problem, a requesting process eventually executes the critical section but no two requesting processes must execute the critical section concurrently.

We now introduce the notions of {\em abstract configuration, state-projection, and se\-quen\-ce-pro\-jec\-tion}. These three notions are useful to formalize {\em safety-distributed} specifications.  

\begin{definition}[Abstract Configuration]
We call {\em abstract configuration} any configuration restricted to the state of the processes ({\em i.e.}, a configuration where the state of each link has been removed).
\end{definition}

\begin{definition}[State-Projection]
Let $\gamma$ be configuration and $p$ be a process. The {\em state-projection} of $\gamma$ on $p$, noted $\phi_p(\gamma)$, is the local state of $p$ in $\gamma$. Similary, the {\em state-projection} of $\gamma$ on all processes, $\phi(\gamma)$ is the product of the local states of all processes in $\gamma$ ({\em n.b.} $\phi(\gamma)$ is an abstract configuration).
\end{definition}

\begin{definition}[Sequence-Projection]
Let $s = \gamma_0$,$\gamma_1$, \dots be a configuration sequence and $p$ be a process. The {\em sequence-projection} of $s$ on $p$, noted $\Phi_p(s)$, is the state sequence $\phi_p(\gamma_0)$,$\phi_p(\gamma_1)$, \dots\ Similary, the {\em sequence-projection} of $s$ on all processes, noted $\Phi(s)$, is the abstract configuration sequence $\phi(\gamma_0)$,$\phi(\gamma_1)$, \dots
\end{definition}

\begin{definition}[Safety-Distributed]\label{def:SafeD}
A specification $\mathcal{SP}$ is {\em safety-distributed} if there exists a sequence of abstract configurations $\mathtt{BAD}$, called {\em bad-factor}, such that:
\BEGLIST
\item[(1)] For each execution $e$, if there exist three configuration sequences $e_0$, $e_1$, and $e_2$ such that $e=e_0e_1e_2$ and $\Phi(e_1) = \mathtt{BAD}$, then $e$ does not satisfy $\mathcal{SP}$.
\item[(2)] For each process $p$, there exists at least one execution $e_p$ satisfying $\mathcal{SP}$ where there exist three configuration sequences $e_p^0$, $e_p^1$, and $e_p^2$ such that $e_p = e_p^0e_p^1e_p^2$ and $\Phi_p(e_p^1) = \Phi_p(\mathtt{BAD})$. 
\ENDLIST
\end{definition}

\noindent Almost all classical problems of distributed computing have {\em safety-distributed} specifications, {\em e.g.}, mutual exclusion, phase synchronization, \dots\ For example, in mutual exclusion a {\em bad-factor} is any sequence of abstract configurations where several requesting processes executes the critical section concurrently. We now consider a message-passing system with unbounded capacity channels and show the impossibility of {\em snap-stabilization} for {\em safety-distributed} specifications in that case.  

\begin{theorem}\label{theo:infty}
There exists no {\em safety-distributed} specification that admits a snap-stabilizing solution in message-passing systems with unbounded capacity channels. 
\end{theorem}
\begin{proof}
 Let $\mathcal{SP}$ be a {\em safety-distributed} specification and $\mathtt{BAD} = \alpha_0$,$\alpha_1$,\dots be a {\em bad-factor} of $\mathcal{SP}$. 

Assume, for the purpose of contradiction, that there exists a protocol $\mathcal{P}$ that is snap-stabilizing for $\mathcal{SP}$. By Definition \ref{def:SafeD}, for each process $p$, there exists an execution $e_p$ of $\mathcal{P}$ that can be split into three execution factors $e_p^0$, $e_p^1 = \beta_0$,$\beta_1$,\dots, and $e_p^2$ such that $e_p = e_p^0e_p^1e_p^2$ and $\Phi_p(e_p^1) = \Phi_p(\mathtt{BAD})$. Let us denote by $MesSeq_p^q$ the ordered sequence of messages that $p$ receives from any process $q$ in $e_p^1$. Consider now the configuration $\gamma_0$ such that:
\BEGLIST
\item[(1)] $\phi(\gamma_0) = \alpha_0$.
\item[(2)] For each two processes $p$, $q$ such that $p \neq q$, the link $\{p$,$q\}$ as the following state in $\gamma_0$:
\BEGLIST
\item[(a)] The messages in the channel from $q$ to $p$ are exactly the sequence $MesSeq_p^q$ (keeping the same order).
\item[(b)] The messages in the channel from $p$ to $q$ are exactly the sequence $MesSeq_q^p$ (keeping the same order).
\ENDLIST
\ENDLIST
(It is important to note that we have the guarantee that $\gamma_0$ exists because we assume unbounded capacity channels. Assuming channels with a bounded capacity $c$, no configuration satisfies Point $(2)$ if there are at least two distinct processes $p$ and $q$ such that $|MesSeq_p^q| > c$.)

As $\mathcal{P}$ is snap-stabilizing, $\gamma_0$ is a possible initial configuration of $\mathcal{P}$. To obtain the contradiction, we now show that there is an execution starting from $\gamma_0$ that does not satisfy $\mathcal{SP}$. By definition, $\phi(\gamma_0) = \alpha_0$. Consider a process $p$ and the two first configurations of $e_p^1$: $\beta_0$ and $\beta_1$. Any message that $p$ receives in $\beta_0 \mapsto \beta_1$ can be received by $p$ in the first step from $\gamma_0$: $\gamma_0 \mapsto \gamma_1$. Now, $\phi_p(\gamma_0) = \phi_p(\beta_0)$. So, $p$ can behave in $\gamma_0 \mapsto \gamma_1$ as in $\beta_0 \mapsto \beta_1$. In that case, $\phi_p(\gamma_1) = \phi_p(\beta_1)$. Hence, if every process $p$ behaves in $\gamma_0 \mapsto \gamma_1$ as in the first step of its execution factor $e_p^1$, we obtain a configuration $\gamma_1$ such that $\phi(\gamma_1) = \alpha_1$. By induction principle, there exists an execution prefix starting from $\gamma_0$ noted $PRED$ such that $\Phi(PRED) = \mathtt{BAD}$. As $\mathcal{P}$ is snap-stabilizing, there exists an execution $SUFF$ that starts from the last configuration of $PRED$. Now, merging $PRED$ and $SUFF$ we obtain an execution of $\mathcal P$ that does not satisfy $\mathcal{SP}$ --- this contradicts the fact that $\mathcal P$ is snap-stabilizing. 
\end{proof}

\noindent Intuitively, the impossibility result of Theorem \ref{theo:infty} is due to the fact that  in a system with unbounded capacity channels, any initial configuration can contain an unbounded number of messages. If we consider now systems with bounded and known channel capacity, we can circumvent the impossibility result by designing protocols that require a number of messages that is greater than the bound on the channel capacity to perform their specified task. This is our approach in the next section.

%% file: pif.tex
\subsection{A PIF Protocol}

The concept of {\em Propagation of Information with Feedback} (PIF), also called {\em Wave Propagation}, has been introduced by Chang \cite{C82} and Segall \cite{Seg83}. PIF has been extensively studied in the distributed literature because many fundamental protocols, {\em e.g.}, {\em Reset}, {\em Snapshot}, {\em Leader Election}, and  {\em Termination Detection}, can be solved using a PIF-based solution. The PIF scheme can be informally described as follows: when requested, a process starts the first phase of the PIF-computation by broadcasting a specific message {\em m} into the network (this phase is called the {\em broadcast phase}). Then, every non-initiator acknowledges\footnote{An acknowledgment is a message sent by the receiving process to inform the sender about data it have correctly received.} to the initiator the receipt of {\em m} (this phase is called the {\em feedback phase}). The PIF-computation terminates when the initiator received acknowledgments from every other process and decides taking these acknowledgments into account. In distributed systems, any process may need to initiate a PIF-computation. Thus, any process can be the initiator of a PIF-computation and several PIF-computations may run concurrently. Hence, any PIF protocol has to cope with concurrent PIF-computations.

\begin{specification}[PIF-Execution]\label{spec:pif} An execution $e$ satisfies {\em PIF-execution}($e$) \IFF $e$ satisfies the following four properties:
\BEGLIST
\item[-] {\bf Start.} When there is a request for a process $p$ to broadcast a message {\em m}, $p$ starts a {\em PIF-computation} in finite time. 
\item[-] {\bf Correctness.} During any {\em PIF-computation} started by $p$ for the message {\em m}: 
\BEGLIST
\item[-] Any process different of $p$ receives {\em m}.
\item[-] $p$ receives acknowledgments for {\em m} from every other process.
\ENDLIST
\item[-] {\bf Termination.} Any PIF-computation (even non-started) terminates in finite time. 
\item[-] {\bf Decision.} When a {\em PIF-computation} started by $p$ terminates at $p$, $p$ decides taking all acknowledgments of the last message it broadcasts into account only.
\ENDLIST
\end{specification}

\paragraph{Approach.} 

In the following, we refer to our snap-stabilizing PIF as Protocol $\pif$. We describe our approach using a network of two processes: $p$ and $q$. The generalization to a fully-connected network of more than two processes is straightforward and presented in Algorithm \ref{algo:PIF}. 

Consider the following example. Each process maintains in the variable $Old$ its own age and $p$ wants to know the age of $q$. Then, $p$ performs a PIF of the message ``How old are you?''. To that goal, we need the following input\slash output variables:
\BEGLIST
\item[-] {\em $\req_p$}. This variable is used to manage the PIF'requests for $p$. $\req_p$ is (externally) set to $\rwait$ when there is a request for $p$ to perform a PIF. $\req_p$ is switched from $\rwait$ to $\rin$ at the start of each PIF-computation ({\em n.b.} $p$ starts a PIF-computation upon a request only). Finally, $\req_p$ is switched from $\rin$ to $\rdone$ at the termination of each PIF-computation (this latter switch also corresponds to the {\em decision event}). Since a PIF-computation is started by $p$, we assume that $p$ does not set $\req_p$ to $\rwait$ until the termination of the current PIF-computation, {\em i.e.}, until $\req_p = \rdone$.
\item[-] {\em $\mes_p$}. This variable contains the message to broadcast.
\item[-] {\em $\fmes_q$}. When $q$ receives the broadcast message, $q$ assigns the acknowledgment message in $\fmes_q$. 
\ENDLIST
\noindent Using these variables, we perform a PIF of ``How old are you?'' as follows: $\pif .\mes_p$ and $\pif .\req_p$ are respectively (externally) set to ``How old are you?'' and $\rwait$ meaning that we request that $p$ broadcasts ``How old are you?'' to $q$. Consequently to this request, Protocol $\pif$ starts a PIF-computation by setting $\pif .\req_p$ to $\rin$ and this computation terminates when $\pif .\req_p$ is set to $\rdone$. Between this start and this termination, $\pif$ generates two events. First, a ``\BRD$\langle How\ old\ are\ you? \rangle$ \FROM $p$'' event at $q$. When this event occurs, $q$ sets $\pif .\fmes_q$ to $Old_q$ so that $\pif$ feedbacks the value of $Old_q$ to $p$. Protocol $\pif$ then transmits the value of $Old_q$ to $p$: this generates a ``\FCK$\langle x \rangle$ \FROM $q$'' event at $p$ where $x$ is the value of $Old_q$.

A naive attempt to implement Protocol $\pif$ could be the following: 
\BEGLIST
\item[-] When $\pif .\req_p = \rwait$, $p$ sends a broadcast message containing the data message $\pif .\mes_p$ to $q$ and sets $\pif .\req_p$ to $\rin$ (meaning that the PIF-computation is in processing). 
\item[-] Upon receiving a broadcast message containing the data $B$, a ``\BRD$\langle B \rangle$ \FROM $p$'' event is generated at $q$ so that the application (at $q$) that uses the PIF treats the message $B$. Upon this event, the application is assumed to set the feedback message into $\pif .\fmes_q$. Then, $q$ sends a feedback message containing $\pif.\fmes_q$ to $p$.
\item[-] Upon receiving a feedback message containing the data $F$, a ``\FCK$\langle F \rangle$ \FROM $q$'' event is generated at $p$ so that the application (at $p$) that uses the PIF treats the feedback and then sets $\pif .\req_p$ to $\rdone$.
\ENDLIST

\noindent Unfortunately, such a simple approach is not snap-stabilizing in our system:
\BEGLIST
\item[(1)] Due to the unreliability of the channels, the system may suffer of deadlock. If the broadcast message from $p$ or feedback message from $q$ are lost, Protocol $\pif$ never terminates at $p$.
\item[(2)] Due to the arbitrary initial configuration, the link $\{p$,$q\}$ may initially already contain an arbitrary message in the channel from $p$ to $q$ and another in the channel from $q$ to $p$. Hence, after sending the broadcast message to $q$, $p$ may receive a feedback message that was not sent by $q$. Also, $q$ may receive a broadcast message that was not sent by $p$: as a consequence, $q$ generates an undesirable feedback message.
\ENDLIST

\noindent To circumvent these two problems, we use two additionnal variables at each process:
\BEGLIST
\item[-] $\state_p \in \{0$,$1$,$2$,$3$,$4\}$ (resp. $\state_q$) is a flag value that $p$ (resp. $q$) puts into its messages. 
\item[-] $\neigstate_p$ (resp. $\neigstate_q$) is equal to the last $\state_q$ (resp. $\state_p$) that $p$ (resp. $q$) receives from $q$ (resp. $p$).
\ENDLIST

\noindent (Note that we use a single message type, noted $\pifm$, to manage the PIF-computations initiated by both $p$ and $q$.)

Our protocol works as follows: $p$ starts a {\em PIF-computation} by setting $\state_p$ to $0$. Then,  until $\state_p$ $=$ $4$, $p$ repeatedly sends $\langle \pifm$,$\mes_p$,$\fmes_p$,$\state_p$,$\neigstate_p \rangle$ to $q$. When $q$ receives $\langle B$,$F$,$pState$,$qState \rangle$ (from $p$), $q$ updates $\neigstate_q$ to $pState$ and then sends a message $\langle \pifm$,$\mes_q$,$\fmes_q$,$\state_q$,$\neigstate_q \rangle$ to $p$ if $pState < 4$ ({\em i.e.}, if $p$ is still waiting for a message from $q$). Finally, $p$ increments $\state_p$ only when it receives a $\langle \pifm$,$B$,$B$,$qState$,$pState \rangle$ message from $q$ such that $\state_p = pState$ and $pState < 4$. Hence, after $p$ starts, $\state_p = 4$ only after $p$ successively receives $\langle \pifm$,$B$,$F$,$qState$,$pState \rangle$ messages (from $q$) with $pState = 0$,$1$,$2$,$3$. Now, considering the arbitrary initial value of $\neigstate_q$ and the at most two arbitrary messages initially in the link $\{p$,$q\}$ (one in the channel from $p$ to $q$ and one in the channel from $q$ to $p$), we are sure that after $p$ starts, $p$ receives a $\langle \pifm$,$B$,$F$,$qState$,$pState \rangle$ from $q$ with $pState = State_p = 3$ only if this message was sent by $q$ consequently to the reception by $q$ of a message sent by $p$. 

\begin{figure*}[htpb]
\begin{center}
\includegraphics[width=250pt,height=60pt]{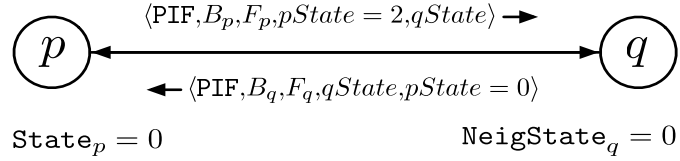}
\caption{Worst case of Protocol $\pif$ in terms of configurations.\label{ex1}}
\end{center}
\end{figure*}

Figure \ref{ex1} illustrates the worst case of Protocol $\pif$ in terms of configurations. In this example, $p$ may increment $\state_p$ after receiving the initial message with the flag value $pState = 0$. Then, if $q$ starts a PIF-computation, $q$ sends messages with the flag value $pState = 1$ until receiving (from $p$) the initial message with the value $pState = 2$. Hence, $p$ can still increment $\state_p$ twice due to the values 1 and 2 ({\em i.e.}, $\state_p$ then reaches the value 3). But, after these incrementations, $p$ no more increments $\state_p$ until receiving a message with the value $pState = 3$ and $q$ starts sending messages with the value $pState = 3$ only after receiving a message from $p$ with the value $pState = 3$. Finally, note that after receiving a message with the value $pState = 3$, $p$ increments $\state_p$ to 4 and stops sending messages until the next request. This ensures that if the requests eventually stop, the system eventually contains no message.

\smallskip

\noindent It remains to see when a process can generate the \BRD\ and \FCK\ events:
\BEGLIST 
\item[-] $q$ receives at least 4 copies of the broadcast messages. But, $q$ generate a \BRD\ event only once for each broadcast message: when $q$ switches $\neigstate_q$ to 3.
\item[-] After it starts, $p$ is sure to receive the ``good'' feedback only when it receives a message with $pState = \state_p = 3$. As previously, to limit the number of events, $p$ generates a \FCK\ events only when it switches $\state_p$ from 3 to 4. The other copies are then ignored. Also, note that after receiving this message, $p$ can only receives duplicates until the next PIF-computation. Hence, when $p$ decides, it decides only taking the ``good'' feedbacks into account.
\ENDLIST

\noindent We generalize this snap-stabilizing one-to-one broadcast with feedback to a snap-stabilizing all-to-all broadcast with feedback ({\em i.e.}, a PIF) in Algorithm \ref{algo:PIF}. It is important to note that our protocol does not prevent processes to generate unexpected \BRD\ or \FCK\ events. Actually, what our protocol ensures is: when a process $p$ starts to broadcast a message {\em m}, then (1) every other process eventually receives {\em m} (\BRD), (2) $p$ eventually receives a feedback for {\em m} from any other process (\FCK), and (3) $p$ decides ($\pif .\req_p \gets \rdone$) by only taking the ``good'' feedbacks into account. Another interesting property of our protocol is the following: after the first complete computation of $\pif$ (from the start to the termination), the channels from and to $p$ contain no message from the initial configuration.  

\begin{algorithm}[t]\caption{Protocol $\pif$ for any process $p$\label{algo:PIF}}
\scriptsize

\smallskip

\noindent {\bf Constant:} $n$: integer, number of processes

\smallskip

\noindent {\bf Variables:}

\smallskip

\begin{tabular}{lll}
$\req_p \in \{\rwait$,$\rin$,$\rdone\}$             & : & input\slash output variable\\
$\mes_p$                                            & : & data to broadcast, input variable\\
$\fmes_p[1\dots n-1]$                               & : & array of messages to feedback, input variable\\
$\state_p[1\dots n-1] \in \{0$,$1$,$2$,$3$,$4\}^{n-1}$   & : & internal variable\\   
$\neigstate_p[1\dots n-1] \in \{0$,$1$,$2$,$3$,$4\}^{n-1}$  & : & internal variable            
\end{tabular}

\smallskip

\noindent {\bf Actions:}

\smallskip

\begin{tabular}{lll}
\noindent $\rone$ $::$ $(\req_p = \rwait)$ & $\to$ & $\req_p \gets \rin$ \Rem{Start}\\
                 &       & \FOR $q \in [1\dots {n-1}]$ \DO\\
                 &       & \qquad $\state_p[q] \gets 0$\\
                 &       & \DONE\\
\\
\noindent $\rtwo$ $::$ $(\req_p = \rin)$   & $\to$ & \IF ($\forall q \in [1\dots {n-1}]$, $\state_p[q] = 4$) \THEN\\
                 &       & \qquad $\req_p \gets \rdone$ \Rem{Termination}\\
                 &       & \ELSE\\
                 &       & \qquad \FOR $q \in [1\dots {n-1}]$ \DO\\
                 &       & \qquad \qquad \IF $(\state_p[q] \neq 4)$ \THEN\\
                 &       & \qquad \qquad \qquad \SEND$\langle$$\pifm$,$\mes_p$,$\fmes_p[q]$,$\state_p[q]$,$\neigstate_p[q]$$\rangle$\TO $q$\\
                 &       & \qquad \qquad \FI\\
                 &       & \qquad \DONE\\
                 &       & \FI\\
\\
\noindent $\rthree$ $::$ \RECV$\langle$$\pifm$,$B$,$F$,$qState$,$pState$$\rangle$\FROM $q$ & $\to$ & \IF $(\neigstate_p[q] \neq 3) \wedge (qState = 3)$ \THEN\\ 
                 &       & \qquad generate a ``\BRD$\langle B \rangle$ \FROM $q$'' event\\
                 &       & \FI\\
                 &       & $\neigstate_p[q] \gets qState$\\
                 &       & \IF $(\state_p[q] = pState) \wedge (\state_p[q] < 4)$ \THEN\\
                 &       & \qquad $\state_p[q] \gets \state_p[q] + 1$\\
                 &       & \qquad \IF ($\state_p[q] = 4$) \THEN\\
                 &       & \qquad \qquad generate a ``\FCK$\langle F \rangle$ \FROM $q$'' event\\
                 &       & \qquad \FI\\
                 &       & \FI\\
                 &       & \IF $(qState < 4)$ \THEN\\ 
                 &       & \qquad \SEND$\langle$$\pifm$,$\mes_p$,$\fmes_p[q]$,$\state_p[q]$,$\neigstate_p[q]$$\rangle$\TO $q$\\
                 &       & \FI\\
\end{tabular}
\end{algorithm}

\paragraph{Proof of Snap-Stabilization.}

\noindent The proof of snap-stabilization of $\pif$ just consists in showing that, despite the arbitrary initial configuration, any execution of $\pif$ always satisfies the four properties of Specification \ref{spec:pif}. In the following proofs, the $message\mbox{-}values$ will be replaced by ``$-$'' when they have no impact on the reasonning.

\begin{lemma}[Start]\label{lem:pif:start}
Starting from any configuration, when there is a request for a process $p$ to broadcast a message, $p$ starts a {\em PIF-computation} in finite time. 
\end{lemma}
\begin{proof}
We assumed that $\req_p$ is externally set to $\rwait$ when there is a request for the process $p$ to broadcast a message. Moreover, we claim that a process $p$ starts Protocol $\pif$ by switching $\req_p$ from $\rwait$ to $\rin$. Now, when $\req_p = \rwait$, Action $\rone$ is continuously enabled at $p$ and by executing $\rone$, $p$ sets $\req_p$ to $\rin$. Hence, the lemma holds.
\end{proof}

\noindent The following Lemmas (Lemmas \ref{lem:incr}-\ref{lem:pif:dec}) hold assuming that no {\em PIF-computation} (even non-started) can be interrupted due to another request:

\begin{hypothesis}\label{hyp:request}
While $\req_p \neq \rdone$, $\req_p$ is not (externally) set to $\rwait$.
\end{hypothesis}

\begin{lemma}\label{lem:incr}
Consider two distinct processes $p$ and $q$. Starting from any configuration, if $(\req_p$ $=$ $\rin) \wedge (\state_p[q] < 4)$, then $\state_p[q]$ is eventually incremented.
\end{lemma}
\begin{proof}
Assume, for the purpose of contradiction, that $\req_p = \rin$ and $\state_p[q] = i$ with $i < 4$ but $\state_p[q]$ is never incremented. Then, from Algorithm \ref{algo:PIF}, $\req_p = \rin$ and $\state_p[q] = i$ hold forever and by Actions $\rtwo$ and $\rthree$, we know that:
\BEGLIST
\item[-] $p$ only sends to $q$ messages of the form $\langle \pifm$,$-$,$-$,$i$,$- \rangle$.
\item[-] $p$ sends such messages infinitely many times.
\ENDLIST
As a consequence, $q$ eventually only receives from $p$ messages of the form $\langle \pifm$,$-$,$-$,$i$,$- \rangle$ and $q$ receives such messages infinitely often. By Action $\rthree$, $\neigstate_q[p] = i$ eventually holds forever. From that point, any message that $q$ sends to $p$ is of the form $\langle \pifm$,$-$,$-$,$-$,$i \rangle$. Also, as $i < 4$ and $q$ receives infinitely many messages from $p$, $q$ sends infinitely many messages of the form $\langle \pifm$,$-$,$-$,$-$,$i \rangle$ to $p$ (see Action $\rthree$). Hence, $p$ eventually receives $\langle \pifm$,$-$,$-$,$-$,$i \rangle$ from $q$ and, as a consequence, increments $\state_p[q]$ (see Action $\rthree$) --- a contradiction.
\end{proof}

\begin{lemma}[Termination]\label{lem:pif:term}
Starting from any configuration, any {\em PIF-computation} (even non-started) terminates in finite time. 
\end{lemma}
\begin{proof}
Assume, for the purpose of contradiction, that a {\em PIF-computation} never terminates at some process $p$, {\em i.e.}, $\req_p \neq \rdone$ forever. Then, $\req_p = \rin$ eventually holds forever by Lemma \ref{lem:pif:start}. Now, by Lemma \ref{lem:incr} and owing the fact that $\forall q \in [1\dots n-1]$, $\state_p[q]$ cannot decrease while the computation is not terminated at $p$, we can deduce that $p$ eventually satisfies ``$\forall q \in [1\dots n-1]$,$\state_p[q] = 4$'' forever. In this case, $p$ sets $\req_p$ to $\rdone$ by Action $\rtwo$ --- a contradiction. 
\end{proof}

\begin{lemma}\label{lem:pif:23}Let $p$ and $q$ be two distinct processes. After $p$ starts to broadcast a message from an arbitrary configuration, $p$ switches $\state_p[q]$ from 2 to 3 only if the three following conditions hold:
\BEGLIST
\item[(1)] Any message in the channel from $p$ to $q$ are of the form $\langle \pifm$,$-$,$-$,$i$,$- \rangle$ with $i \neq 3$.
\item[(2)] $\neigstate_q[p] \neq 3$.
\item[(3)] Any message in the channel from $q$ to $p$ are of the form $\langle \pifm$,$-$,$-$,$-$,$j \rangle$ with $j \neq 3$.
\ENDLIST
\end{lemma}
\begin{proof} $p$ starts to broadcast a message by executing Action $\rone$ ({\em n.b.} $\rone$ is the only {\em starting action} of $\pif$). When $p$ executes $\rone$, $p$ sets (in particular) $\state_p[q]$ to 0. From that point, $\state_p[q]$ can only be incremented one by one until reaching value 4. Let us study the three first incrementations of $\state_p[q]$:
\BEGLIST
\item[-] {\bf From 0 to 1.} $\state_p[q]$ switches from 0 to 1 only after $p$ receives $\langle \pifm$,$-$,$-$,$-$,$0 \rangle$ from $q$ (Action $\rthree$). As the link $\{p$,$q\}$ always contains at most one message in the channel from $q$ to $p$, the next message that $p$ will receive from $q$ will be a message sent by $q$.
\item[-] {\bf From 1 to 2.} From the previous case, we know that $\state_p[q]$ switches from 1 to 2 only when $p$ receives $\langle \pifm$,$-$,$-$,$-$,$1 \rangle$ from $q$ and this message was sent by $q$. From Actions $\rtwo$ and $\rthree$, we can then deduce that $\neigstate_q[p] = 1$ held when $q$ sent $\langle \pifm$,$-$,$-$,$-$,$1 \rangle$ to $p$. From that point, $\neigstate_q[p] = 1$ holds until $q$ receives from $p$ a message of the form $\langle \pifm$,$-$,$-$,$i$,$- \rangle$ with $i \neq 1$.
\item[-] {\bf From 2 to 3.} The switching of $\state_p[q]$ from 2 to 3 can occurs only after $p$ receives a message $mes_1 = \langle \pifm$,$-$,$-$,$-$,$2 \rangle$ from $q$. Now, from the previous case, we can deduce that $p$ receives $mes_1$ consequently to the reception by $q$ of a message $mes_0 = \langle \pifm$,$-$,$-$,2,$- \rangle$ from $p$. Now:
\BEGLIST
\item[(a)] As the link $\{p$,$q\}$ always contains at most one message in the channel from $p$ to $q$, after receiving $mes_0$ and until $\state_p[q]$ switches from 2 to 3, every message in transit from $p$ to $q$ is of the form $\langle \pifm$,$-$,$-$,$i$,$- \rangle$ with $i \neq 3$ (Condition $(1)$ of the lemma) because after $p$ starts to broadcast a message, $p$ sends messages of the form $\langle \pifm$,$-$,$-$,$3$,$- \rangle$ to $q$ only when $\state_p[q] = 3$.
\item[(b)]  After receiving $mes_0$, $\neigstate_q[p] \neq 3$ until $q$ receives $\langle \pifm$,$-$,$-$,3,$- \rangle$. Hence, by $(a)$, after receiving $mes_0$ and until (at least) $\state_p[q]$ switches from 2 to 3, $\neigstate_q[p] \neq 3$  (Condition $(2)$ of the lemma).
\item[(c)] After receiving $mes_1$, $\state_p[q] \neq 3$ until $p$ receives $\langle \pifm$,$-$,$-$,$-$,$3 \rangle$ from $q$. As $p$ receives $mes_1$ after $q$ receives $mes_0$, by $(b)$ we can deduce that after receiving $mes_1$ and until (at least) $\state_p[q]$ switches from 2 to 3, every message in transit from $q$ to $p$ is of the form $\langle \pifm$,$-$,$-$,$-$,$j \rangle$ with $j \neq 3$ (Condition $(3)$ of the lemma).
\ENDLIST
Hence, when $p$ switches $\state_p[q]$ from 2 to 3, the three conditions $(1)$, $(2)$, and $(3)$ are satisfied, which proves the lemma.
\ENDLIST  
\end{proof}

\begin{lemma}[Correctness]\label{lem:pif:correct}
Starting from any configuration, if $p$ starts to broadcast a message {\em m}, then: 
\BEGLIST
\item[-] Any process different of $p$ receives {\em m}.
\item[-] $p$ receives acknowledgments for {\em m} from every other process.
\ENDLIST
\end{lemma}
\begin{proof}
$p$ starts to broadcast {\em m} by executing Action $\rone$: $p$ switches $\req_p$ from $\rwait$ to $\rin$ and sets $\state_p[q]$ to 0, $\forall q \in [1\dots 0]$. Then, $\req_p$ remains equal to $\rin$ until $p$ decides by $\req_p \gets \rdone$. Now, $p$ decides in finite time by Lemma \ref{lem:pif:term} and when $p$ decides, we have $\state_p[q] = 4$, $\forall q \in [1\dots 0]$ (Action $\rtwo$). From the code of Algorithm \ref{algo:PIF}, this means that $\forall q \in [1\dots 0]$, $\state_p[q]$ is incremented one by one from 0 to 4. By Lemma \ref{lem:pif:23}, $\forall q \in [1\dots 0]$, $\state_p[q]$ is incremented from 3 to 4 only after:
\BEGLIST
\item[-] $q$ receives a message sent by $p$ of the form $\langle \pifm$,{\em m},$-$,3,$- \rangle$, and then
\item[-] $p$ receives a message sent by $q$ of the form $\langle \pifm$,$-$,$-$,3,$- \rangle$.  
\ENDLIST
When $q$ receives the first $\langle \pifm$,{\em m},$-$,3,$- \rangle$ message from $p$, $q$ generates a ``\BRD$\langle${\em m}$\rangle$ \FROM $p$'' event and then starts to send $\langle \pifm$,$-$,$F$,$-$,$3 \rangle$ messages to $p$\footnote{$q$ sends a $\langle \pifm$,$-$,$F$,$-$,$3 \rangle$ message to $p$ (at least) each time it receives a $\langle \pifm$,{\em m},$-$,3,$- \rangle$ message from $p$.}. From that point and until $p$ decides, $q$ only receives $\langle \pifm$,{\em m},$-$,3,$- \rangle$ message from $p$. So, from that point and until $p$ decides, any message that $q$ sends to $p$ acknowledges the reception of {\em m}. Since, $p$ receives the first $\langle \pifm$,$-$,$F$,$-$,$3 \rangle$ message from $q$, $p$ generates a ``\FCK$\langle F\rangle$ \FROM $q$'' event and then sets $\state_p[q]$ to 4.

Hence, $\forall q \in [1\dots 0]$, the broadcast of {\em m} generates a ``\BRD$\langle${\em m}$\rangle$ \FROM $p$'' event at process $q$ and then an associated ``\FCK$\langle F\rangle$ \FROM $q$'' event at $p$, which proves the lemma.
\end{proof}

\begin{lemma}[Decision]\label{lem:pif:dec} 
Starting from any configuration, when a {\em PIF-computation} started by $p$ terminates at $p$, $p$ decides taking all acknowledgments of the last message it broadcasts into account only.
\end{lemma}
\begin{proof} First, $p$ starts to broadcast a message {\em m} by executing Action $\rone$: $p$ switches $\req_p$ from $\rwait$ to $\rin$ and sets $\state_p[q]$ to 0, $\forall q \in [1\dots 0]$. Then, $\req_p$ remains equal to $\rin$ until $p$ decides by $\req_p \gets \rdone$. Now, $(1)$ $p$ decides in finite time by Lemma \ref{lem:pif:term}, $(2)$ when $p$ decides, we have $\state_p[q] = 4$, $\forall q \in [1\dots 0]$ (Action $\rtwo$), and $(3)$ after $p$ decides, each time $q$ receives a message from $p$ with the data {\em m}, the message is ignored (this is a consequence of Claim $(2)$). From the code of Algorithm \ref{algo:PIF}, we know that exactly one ``\FCK$\langle F\rangle$ \FROM $q$'' event per neighbor $q$ occurs at $p$ before $p$ decides: when $p$ switches $\state_p[q]$ from 3 to 4. Now, Lemma \ref{lem:pif:correct} and Claim $(3)$ imply that each of these feedbacks corresponds to an acknowledgment for {\em m}. Hence, $p$ decides taking all acknowledgments of {\em m} into account only and the lemma is proven.
\end{proof}

\noindent By Lemmas \ref{lem:pif:start}, \ref{lem:pif:term}, \ref{lem:pif:correct}, and \ref{lem:pif:dec}, starting from any arbitrary initial configuration, any execution of $\pif$ always satisfies Specification \ref{spec:pif}. Hence, follows:

\begin{theorem}
Protocol $\pif$ is snap-stabilizing for Specification \ref{spec:pif}.
\end{theorem}

\noindent Below, we give an additionnal property of $\pif$, this property will be used in the snap-stabilization proof of Protocol $\me$.

\begin{property}\label{prop:pif:flush}
If $p$ starts a PIF-computation (using Protocol $\pif$) in the configuration $\gamma_0$ and the computation terminates at $p$ in the configuration $\gamma_k$, then any message that was in a channel from and to $p$ in $\gamma_0$ is no longer in the channel in $\gamma_k$.
\end{property}
\begin{proof}
Assume that a process $p$ starts a PIF-computation (using Protocol $\pif$) in the configuration $\gamma_0$. Then, as $\pif$ is snap-stabilizing for Specification \ref{spec:pif}, we have the guarantee that for every $p$'neighbor $q$, at least one broadcast message crosses the channel from $p$ to $q$ and at least one acknowledgment message crosses the channel from $q$ to $p$ during the PIF-computation. Now, we assumed that each channel has a single-message capacity. Hence, every message that was in a channel from and to $p$ in the configuration $\gamma_0$ has been received or lost when the PIF-computation terminates at $p$ in configuration $\gamma_k$ 
\end{proof}

%% file: id.tex
\subsection{A IDs-Learning Protocol}

Protocol $\IDL$ (its implementation is presented in Algorithm \ref{algo:IDL}) is a simple application of Protocol $\pif$. This protocol assumes IDs on processes ($ID_p$ denotes the identity of the process $p$) and uses three variables at each process $p$:
\BEGLIST
\item[-] {\em $\req_p \in \{\rwait$,$\rin$,$\rdone\}$}. The goal of this variable is the same as in $\pif$.
\item[-] {\em $\Min_p$}. After a complete execution of $\IDL$ ({\em i.e.}, from the start to the termination), $\Min_p$ contains the minimal ID of the system.
\item[-] {\em $\IDtab_p[1\dots n]$}. After a complete execution of $\IDL$, $\IDtab_p[q]$ contains the ID of the $p$'neighbor $q$.
\ENDLIST
When requested ($\IDL .\req_p$ $=$ $\rwait$) at $p$, Protocol $\IDL$ evaluates the ID of each of its neighbors $q$ and the minimal ID of the system using Protocol $\pif$. The results of the computation are available for $p$ since $p$ decides (when $\IDL .\req_p \gets \rdone$). Based on the specification of $\pif$, it is easy to see that $\IDL$ is snap-stabilizing for the following specification:

\begin{specification}[IDs-Learning-Execution]\label{spec:IDL} An execution $e$ satisfies {\em IDs-Lear\-ning-exe\-cu\-tion}($e$) \IFF $e$ satisfies the following four properties:
\BEGLIST
\item[-] {\bf Start.} When requested, a process $p$ starts a {\em IDs-Learning-computation} in finite time. 
\item[-] {\bf Correctness.} At the end of any {\em IDs-Learning-computation} started by $p$: 
\BEGLIST
\item[-] $\forall q \in [1\dots n-1]$, $\IDtab_p[q] = ID_q$.
\item[-] $\Min_p = \min(\{ID_q$, $q \in [1\dots n-1]\} \cup \{ID_p\})$.
\ENDLIST
\item[-] {\bf Termination.} Any {\em IDs-Learning-computation} (even non-started) terminates in finite time. 
\item[-] {\bf Decision.} If $p$ is in a {\em terminal state} and a {\em IDs-Learning-computation} was started by $p$, then $p$ decided knowing the minimal ID of the system and the ID of every of its neighbors.
\ENDLIST
\end{specification}

\begin{theorem}\label{theo:idl:snap}
Protocol $\IDL$ is snap-stabilizing for Specification \ref{spec:IDL}.
\end{theorem}

\begin{algorithm}[htpb]\caption{Protocol $\IDL$ for any process $p$\label{algo:IDL}}
\scriptsize

\smallskip

\noindent {\bf Constant:} 

\smallskip

\begin{tabular}{lll}
$n$    & : & integer, number of processes\\
$ID_p$ & : & integer, identity of $p$\\
\end{tabular}

\smallskip

\noindent {\bf Variables:}

\smallskip

\begin{tabular}{lll}
$\req_p \in \{\rwait$,$\rin$,$\rdone\}$             & : & input\slash output variable\\
$\Min_p$                                            & : & integer, output variable\\
$\IDtab_p[1\dots n-1] \in \N^{n-1}$ & : & output variable\\                              
\end{tabular}

\smallskip

\noindent {\bf Actions:}

\smallskip

\begin{tabular}{lllll}
\noindent $\rone$ & $::$ & $(\req_p = \rwait)$ & $\to$ & $\req_p \gets \rin$ \Rem{Start}\\
&&                 &       & $\Min_p \gets ID_p$\\
&&                 &       & $\pif .\mes_p \gets \midc$\\
&&                 &       & $\pif .\req_p \gets \rwait$\\
\\
\noindent $\rtwo$ & $::$ & $(\req_p = \rin) \wedge (\pif .\req_p = \rdone)$ & $\to$ & $\req_p \gets \rdone$ \Rem{Termination}\\
\\
\noindent $\rthree$ & $::$ & \BRD$\langle \midc \rangle$ \FROM $q$ & $\to$ & $\pif .\fmes_p[q] \gets ID_p$\\
\\
\noindent $\rfour$ & $::$ & \FCK$\langle qID \rangle$ \FROM $q$ & $\to$ & $\IDtab_p[q] \gets qID$\\
&&                 &      & $\Min_p \gets \min (\Min_p$,$qID)$
\end{tabular}
\end{algorithm}

%% file: me.tex
\subsection{A Mutual Exclusion Protocol}

We now consider the problem of {\em mutual exclusion}. Mutual exclusion is a well-known mechanism allowing to allocate a common resource. Indeed, a mutual-exclusion mechanism ensures that a special section of code, called {\em critical section} (noted $\cs$ in the following), can be executed by at most one process at any time. The processes can use their critical section to access to a shared ressource. Generally, this resource corresponds to a set of shared variables in a common store or a shared hardware device ({\em e.g.}, a printer). The first snap-stabilizing implementation of mutual exclusion is presented in \cite{CDV07bis} but in the state model (a stronger model than the message-passing model). In \cite{CDV07bis}, authors adopt the following specification\footnote{This specification was firstly introduced and justified in \cite{CDPV2003}.}:

\begin{specification}[ME-Execution]\label{spec:ME} An execution $e$ satisfies {\em ME-execution}($e$) \IFF $e$ satisfies the following two properties:
\BEGLIST
\item[-] {\bf Start.} Any process that requests the $\cs$ enters in the $\cs$ in finite time. 
\item[-] {\bf Correctness.} If a requesting process enters in the $\cs$, then it executes the $\cs$ alone.
\ENDLIST
\end{specification}

\paragraph{Approach.} We now propose a snap-stabilizing mutual exclusion protocol called Protocol $\me$. The implementation of $\me$ is presented in Algorithm \ref{algo:ME}. As for the previous solutions, Protocol $\me$ uses the input\slash output variable $\req$. A process $p$ (externally) sets $\me .\req_p$ to $\rwait$ when it requests the access to the $\cs$. Process $p$ is then called a {\em requestor} and assumed to not execute $\me .\req_p \gets \rwait$ until $\me .\req_p = \rdone$, {\em i.e.}, until its current request is done.

The main idea of the protocol is the following: we assume IDs on processes and the process with the smallest ID --- called the {\em leader} --- decides using a variable called $\val$ which process can executes the $\cs$. When a process learns that it is authorized to access the $\cs$:
\BEGLIST
\item[(1)] It first ensures that no other process can execute the $\cs$. 
\item[(2)] It then executes the $\cs$ if it wishes.
\item[(3)] Finally, it notifies to the leader that it releases the $\cs$ so that the leader (fairly) authorizes another process to access the $\cs$.
\ENDLIST
\noindent To apply this scheme, $\me$ executes by phases from Phase $0$ to $4$ in such way that each process goes through Phase 0 infinitely often. For each process $p$, $\phase_p$ denotes in which phase process $p$ is. After requesting the $\cs$ ($\me .\req_p \gets \rwait$), a process $p$ can access the $\cs$ only after executing Phase 0. Indeed, $p$ can access to the $\cs$ only if $\me .\req_p = \rin$ and $p$ switches $\me .\req_p$ from $\rwait$ to $\rin$ only when executing Phase 0. Hence, our protocol has just to ensure that after executing its phase 0, a process always executes the $\cs$ alone. Our protocol offers such a guarantee thanks to the five phases described below:
\BEGLIST
\item[-] {\bf Phase 0.} When a process $p$ is in Phase 0, it starts a computation of $\IDL$, sets $\me .\req_p$ to $\rin$ if $\me .\req_p = \rwait$ ({\em i.e.}, if $p$ requests the $\cs$, then the protocol takes this request into account), and finally switches to Phase 1.
\item[-] {\bf Phase 1.} When a process $p$ is in Phase 1, $p$ waits the termination of $\IDL$ to know (1) the ID of each of its neighbors $q$ ($\IDtab_p[q]$) and (2) the leader of the system ($\IDL .\Min_p$), {\em i.e.}, the process with the smallest ID. Then, $p$ starts a PIF of the message $\ask$ to know which is the process authorized to access the $\cs$ and switches to Phase 2. Upon receiving a message $\ask$ from $p$, any process $q$ answers $\yes$ if $\val_q$ is equal to the channel number of $p$ at $q$, $\no$ otherwise. Of course, $p$ will only take the answer of the leader into account.
\item[-] {\bf Phase 2.} When a process $p$ is in Phase 2, it waits the termination of the PIF started in Phase 1. After $\pif$ terminates, the answers of any neighbors $q$ of $p$ are stored in $\nval_p[q]$ and, so, $p$ knows if it is authorized to access the $\cs$. Actually, $p$ is authorized to access the $\cs$ (see $Winner(p)$) if: (1) $p$ is the leader and $\val_p = 0$ or (2) the leader answers $\yes$ to $p$. If $p$ has the authorization to access the $\cs$, $p$ starts a PIF of the message $\exit$. The goal of this message is to force all other processes to restart to Phase 0. This ensures no other process executes the $\cs$ until $p$ notifies to the leader that it releases the $\cs$. Indeed, due to the arbitrary initial configuration, some process $q \neq p$ may believe that it is authorized to execute the $\cs$: if $q$ never starts Phase 0. On the contrary, after restarting to 0, $q$ cannot receive any authorization from the leader until $p$ notifies to the leader that it releases the $\cs$. Finally, $p$ terminates Phase 2 by switching to Phase 3.
\item[-] {\bf Phase 3.} When a process $p$ is in Phase 3, it waits the termination of the last PIF. After $\pif$ terminates, if $p$ is authorized to execute the $\cs$, then: $p$ executes the $\cs$ if $\me .\req_p = \rin$ ({\em i.e.}, if the system took a request of $p$ into account) and then either (1) $p$ is the leader and switches $\val_p$ from 0 to 1 or (2) $p$ is not the leader and starts a PIF of the message $\exitcs$ to notify to the leader that it releases the $\cs$. Upon receiving such a message, the leader increments its variable $\val$ modulus $n + 1$ to authorize another process to access the $\cs$. Finally, $p$ terminates Phase 3 by switching to Phase 4.
\item[-] {\bf Phase 4.} When a process $p$ is in Phase 4, it waits the termination of the last PIF and then switches to Phase 0.
\ENDLIST

\begin{algorithm}[htpb]\caption{Protocol $\me$ for any process $p$\label{algo:ME}}
\scriptsize

\smallskip

\noindent {\bf Constant:} 

\smallskip

\begin{tabular}{lll}
$n$    & : & integer, number of processes\\
$ID_p$ & : & integer, identity of $p$\\
\end{tabular}

\smallskip

\noindent {\bf Variables:}

\smallskip

\begin{tabular}{lll}
$\req_p \in \{\rwait$,$\rin$,$\rdone\}$             & : & input\slash output variable\\                          
$\phase_p \in \{0$,$1$,$2$,$3$,$4\}$                & : & internal variable\\
$\val_p \in \{0\dots n-1\}$ & : & internal variable\\
$\nval_p[1\dots n-1] \in \{true$,$false\}^{n-1}$ & : & internal variable
\end{tabular}

\smallskip

\noindent {\bf Predicate:}

\smallskip

\begin{tabular}{l}
$Winner(p)$ $\equiv$ $(\IDL .\Min_p$$=$$ID_p$$\wedge$$\val_p$$=$$0) \vee (\exists$$q$$\in$$[1\dots n-1]$, $\nval_p[q]$$\wedge$$\IDL .\IDtab_p[q]$$=$$\IDL.\Min_p)$
\end{tabular}

\smallskip

\noindent {\bf Actions:}

\smallskip

\begin{tabular}{lllll}
\noindent $\rzero$ & $::$ & $(\phase_p = 0)$ & $\to$ & $\IDL .\req_p \gets \rwait$\\
                   &      &                  &       & \IF $\req_p = \rwait$ \THEN\\
                   &      &                  &       & \qquad $\req_p \gets \rin$ \Rem{Start}\\
                   &      &                  &       & \FI\\
                   &      &                  &       & $\phase_p \gets \phase_p + 1$\\
\\
\noindent $\rone$  & $::$ & $(\phase_p = 1) \wedge (\IDL .\req_p = \rdone)$ & $\to$ & $\pif .\mes_p \gets \ask$\\
                   &      &                                                 &       & $\pif .\req_p \gets \rwait$\\
                   &      &                  &       & $\phase_p \gets \phase_p + 1$\\
\\
\noindent $\rtwo$  & $::$ & $(\phase_p = 2) \wedge (\pif .\req_p = \rdone)$ & $\to$ & \IF $Winner(p)$ \THEN\\
                   &      &                                                 &       & \qquad $\pif .\mes_p \gets \exit$\\
                   &      &                                                 &       & \qquad $\pif .\req_p \gets \rwait$\\
                   &      &                                                 &       & \FI\\
                   &      &                                                 &       & $\phase_p \gets \phase_p + 1$\\
\\
\noindent $\rthree$  & $::$ & $(\phase_p = 3) \wedge (\pif .\req_p = \rdone)$ & $\to$ & \IF $Winner(p)$ \THEN\\
                     &      &                                                 &       & \qquad \IF $(\req_p = \rin)$ \THEN\\
                     &      &                                                 &       & \qquad \qquad $\cs$\\
                     &      &                                                 &       & \qquad \qquad $\req_p \gets \rdone$ \Rem{Termination}\\
                     &      &                                                 &       & \qquad \FI\\
                     &      &                                                 &       & \qquad \IF $(\IDL .\Min_p = ID_p)$ \THEN\\
                     &      &                                                 &       & \qquad \qquad $\val_p \gets 1$\\
                     &      &                                                 &       & \qquad \ELSE \\
                     &      &                                                 &       & \qquad \qquad $\pif .\mes_p \gets \exitcs$\\
                     &      &                                                 &       & \qquad \qquad $\pif .\req_p \gets \rwait$\\
                     &      &                                                 &       & \qquad \FI\\
                     &      &                                                 &       & \FI\\
                     &      &                                                 &       & $\phase_p \gets \phase_p + 1$\\
\\
\noindent $\rfour$  & $::$ & $(\phase_p = 4) \wedge (\pif .\req_p = \rdone)$ & $\to$ & $\phase_p \gets 0$\\
\\
\noindent $\rfive$ & $::$ & \BRD$\langle \ask \rangle$ \FROM $q$ & $\to$ & \IF $\val_p = q$ \THEN\\
                   &      &                                                               &       & \qquad $\pif .\fmes_p[q] \gets \yes$\\
                   &      &                                                               &       & \ELSE\\
                   &      &                                                               &       & \qquad $\pif .\fmes_p[q] \gets \no$\\
                   &      &                                                               &       & \FI\\
\\
\noindent $\rsix$ & $::$ & \BRD$\langle \exit \rangle$ \FROM $q$ & $\to$ & $\phase_p \gets 0$\\
                  &      &                                       &       & $\pif .\fmes_p[q] \gets \ok$\\
\\
\noindent $\rseven$ & $::$ & \BRD$\langle \exitcs \rangle$ \FROM $q$ & $\to$ & \IF ($\val_p = q$) \THEN\\
                    &      &                                         &       & \qquad $\val_p \gets (\val_p + 1) \bmod (n + 1)$\\
                    &      &                                         &       & \FI\\
                    &      &                                         &       & $\pif .\fmes_p[q] \gets \ok$\\
\\
\noindent $\reight$ & $::$ & \FCK$\langle \yes \rangle$ \FROM $q$ & $\to$ & $\nval_p[q] \gets true$\\
\\
\noindent $\rnine$ & $::$ & \FCK$\langle \no \rangle$ \FROM $q$ & $\to$ & $\nval_p[q] \gets false$\\ 
\\
\noindent $\rten$ & $::$ & \FCK$\langle \ok \rangle$ \FROM $q$ & $\to$ & \Rem{do nothing}
\end{tabular}
\end{algorithm}

\paragraph{Proof of Snap-Stabilization.}

We begin the proof of snap-stabilization of Protocol $\me$ by showing that, despite the arbitrary initial configuration, any execution of $\me$ always satisfies the correctness property of Specification \ref{spec:ME}.

\smallskip

Assume that a process $p$ requests the $\cs$, {\em i.e.}, $\me .\req_p = \rwait$. Then, $p$ cannot enters in the $\cs$ before executing Action $\rzero$, indeed:
\BEGLIST
\item[-] $p$ enters in the $\cs$ only if $\me .\req_p = \rin$, and
\item[-] Action $\rzero$ is the only action of $\me$ allowing $p$ to set $\me .\req_p$ to $\rin$. 
\ENDLIST
\noindent Hence, to show the correctness property of Specification \ref{spec:ME} (Corollary \ref{coro:me:correct}), we have just to prove that, despite the initial configuration, after $p$ executes Action $\rzero$, if $p$ enters in the $\cs$, then it executes the $\cs$ alone (Lemma \ref{lem:me:CSafterA0}).

\begin{lemma}\label{lem:me:0once}Let $p$ be a process. Starting from any configuration, after $p$ executes $\rzero$, if $p$ enters in the $\cs$, then every other process has switches to Phase 0 at least once.
\end{lemma}
\begin{proof}By checking all the actions of Algorithm \ref{algo:ME}, we can remark that after $p$ executes $\rzero$, $p$ must execute the four actions $\rzero$, $\rone$, $\rtwo$, and $\rthree$ successively to enter in the $\cs$ (in $\rthree$). Also, to execute the $\cs$ in Action $\rthree$, $p$ must satisfy the predicate $Winner(p)$. The value of the predicate $Winner(p)$ only depends on (1) the $\IDL$ computation started in $\rzero$ and (2) the PIF of the message $\ask$ started in $\rone$. Now, this two computations are done when $p$ executes $\rtwo$. So, the fact that $p$ satisfies $Winner(p)$ when executing $\rthree$ implies that $p$ also satisfies $Winner(p)$ when executing $\rtwo$. As a consequence, $p$ starts a PIF of the message $\exit$ in $\rtwo$. Now, $p$ executes $\rthree$ only after this PIF terminates. Hence, $p$ executes $\rthree$ only after every other process executes $\rsix$ ({\em i.e.}, the feedback of the message $\exit$): by this action, every other process switches to Phase 0.
\end{proof}

\begin{definition}[Leader] We call {\em Leader} the process of the system with the smallest ID. In the following, this process will be denoted by $\led$.
\end{definition}

\begin{definition}[Favour] We say that the process $p$ {\em favours} the process $q$ if and only if $(p = q \wedge \val_p = 0) \vee (p \neq q \wedge \val_p = q)$.
\end{definition}

\begin{lemma}\label{lem:me:favour}Let $p$ be a process. Starting from any configuration, after $p$ executes $\rzero$, $p$ enters in the $\cs$ only if the {\em leader} favours $p$ until $p$ releases the $\cs$.
\end{lemma}
\begin{proof}
By checking all the actions of Algorithm \ref{algo:ME}, we can remark that after $p$ executes $\rzero$, $p$ must execute the four actions $\rzero$, $\rone$, $\rtwo$, and $\rthree$ successively to enter in the $\cs$ (in $\rthree$). Moreover, $p$ executes a complete $\IDL$\mbox{-}computation between $\rzero$ and $\rone$. So:
\BEGLIST
\item[$(1)$] $\IDL .\Min_p = ID_{\led}$ when $p$ executes $\rthree$ (by Theorem \ref{theo:idl:snap}, $\IDL$ is snap-stabilizing for Specification \ref{spec:IDL}). 
\item[$(2)$] Also, from the configuration where $p$ executes $\rone$, all messages in the channels from and to $p$ have been sent after $\IDL$ starts at $p$ in Action $\rzero$ (Property \ref{prop:pif:flush}, page \pageref{prop:pif:flush}).
\ENDLIST
\noindent Let us now study the two following cases:
\BEGLIST
\item[-] {\em $p = \led$.} In this case, when $p$ executes $\rthree$, $p$ must satisfy $\val_p = \val_{\led} = 0$ to enter in the $\cs$ by $(1)$. This means that $\led$ favours $p$ (actually itself) when $p$ enters in the $\cs$. Morever, as the execution of $\rthree$ is atomic, $\led$ favours $p$ until $p$ releases the $\cs$ and the lemma holds in this case.
\item[-] {\em $p \neq \led$.} In this case, when $p$ executes $\rthree$, $p$ satisfies $\IDL .\Min_p = ID_{\led}$ by $(1)$. So, $p$ executes the $\cs$ only if $\exists q \in [1\dots n-1]$ such that $\IDL .\IDtab_p[q] = ID_{\led} \wedge \nval_p[q] = true$ (see Predicate $Winner(p)$). To that goal, $p$ must receive a feedback message $\yes$ from $\led$ during the PIF of the message $\ask$ started in Action $\rone$. Now, $\led$ sends such a feedback to $p$ only if $\val_{\led} = p$ when the ``$\BRD \langle \ask \rangle$ from $p$'' event occurs at $\led$ (see Action $\rfive$). Also, since $\led$ satisfies $\val_{\led} = p$, $\led$ updates $\val_p$ only after receiving an $\exitcs$ message from $p$ (see Action $\rseven$). Now, by $(2)$, after $\led$ feedbacks $\yes$ to $p$, $\led$ receives an $\exitcs$ message from $p$ only if $p$ broadcasts $\exitcs$ to $\led$ after releasing the $\cs$ (see Action $\rthree$). Hence, $\led$ favours $p$ until $p$ releases the $\cs$ and the lemma holds in this case.
\ENDLIST
\end{proof}
 
\begin{lemma}\label{lem:me:CSafterA0}
Let $p$ be a process. Starting from any configuration, if $p$ enters in the $\cs$ after executing $\rzero$, then it executes the $\cs$ alone.
\end{lemma}
\begin{proof}
Assume, for the purpose of contradiction, that $p$ enters in the $\cs$ after executing $\rzero$ but executes the $\cs$ concurrently with another process $q$. Then, $q$ also executes Action $\rzero$ before executing the $\cs$ by Lemma \ref{lem:me:0once}. By Lemma \ref{lem:me:favour}, we have the two following property:
\BEGLIST
\item[-] $\led$ favours $p$ during the whole period where $p$ executes the $\cs$.
\item[-] $\led$ favours $q$ during the whole period where $q$ executes the $\cs$.
\ENDLIST
\noindent This contradicts the fact that $p$ and $q$ executes the $\cs$ concurrently because $\led$ always favours exactly one process at a time.
\end{proof}

\begin{corollary}[Correctness]\label{coro:me:correct} Starting from any configuration, if a requesting process enters in the $\cs$, then it executes the $\cs$ alone.
\end{corollary}

\noindent We now show that, despite the arbitrary initial configuration, any execution of $\me$ always satisfies the start property of Specification \ref{spec:ME}.

\begin{lemma}\label{lem:me:infty0}Starting from any configuration, every process $p$ switches to Phase 0 infinitely often.
\end{lemma}
\begin{proof}
Consider the two following cases:
\BEGLIST
\item[-] {\em ``$\BRD \langle \exit \rangle$'' events occur at $p$ infinitely often.} Then, each time such an event occurs at $p$, $p$ switches to Phase 0 (see $\rsix$). So, the lemma holds in this case.
\item[-] {\em Only a finite number of ``$\BRD \langle \exit \rangle$'' events occurs at $p$.} In this case, $p$ eventually reaches a configuration from which it no more executes Action $\rsix$. From this configuration, $\phase_p$ can only be incremented modulus 5 and depending of the value of $\phase_p$, we have the following possibilities:
\BEGLIST
\item[-] {\em $\phase_p = 0$.} In this case, $\rzero$ is continuously enabled at $p$. Hence, $p$ eventually sets $\phase_p$ to 1 (see Action $\rzero$).
\item[-] {\em $\phase_p = i$ with $i>0$.} In this case, Action $\mathtt A_i$ is eventually continuously enabled due to the termination property of $\IDL$ and $\pif$. By executing $\mathtt A_i$, $p$ increments $\phase_p$ modulus 5.
\ENDLIST
\noindent Hence, if only a finite number of ``$\BRD \langle \exit \rangle$'' events occurs at $p$, then $\phase_p$ is eventually incremented modulus 5 infinitely often, which proves the lemma in this case.
\ENDLIST
\end{proof}

\begin{lemma}\label{lem:me:incr}Starting from any configuration, $\val_{\led}$ is incremented modulus $n + 1$ infinitely often.
\end{lemma}
\begin{proof}
Assume, for the purpose of contradiction, that $\val_{\led}$ is eventually no more incremented modulus $n + 1$. We can then deduce that $\led$ eventually favours some process $p$ forever. 

In order to prove the contradiction, we first show that (*) {\em assuming that $\led$ favours $p$ forever, only a finite number of ``$\BRD \langle \exit \rangle$'' events occurs at $p$}. To that goal, assume, for the purpose of contradiction, that an infinite number of ``$\BRD \langle \exit \rangle$'' events occurs at $p$. Then, as the number of processes is finite, there is a process $q \neq p$ that broadcasts $\exit$ messages infinitely often. Now, every PIF-computation terminates in finite time (termination property of Specification \ref{spec:pif}, page \pageref{spec:pif}). So, $q$ performs infinitely many PIF of the message $\exit$. In order to start another PIF of the message $\exit$, $q$ must then successively execute Actions $\rzero$, $\rone$, $\rtwo$. Now, when $q$ executes $\rtwo$ after $\rzero$ and $\rone$, $\IDL .\Min_q = ID_{\led}$ and either (1) $q = \led$ and, as $q \neq p$, $\val_{\led} \neq 0$, or (2) ${\led}$ has feedback $\no$ to the PIF of the message $\ask$ started by $q$ because $\val_{\led} = p \neq q$. In both cases, $q$ satisfies $\neg Winner(q)$ and, as a consequence, does not broadcast $\exit$ (see Action $\rthree$). Hence, $q$ eventually stops to broadcast the message $\exit$ --- a contradiction.

Using Property (*), we now show the contradiction. By Lemma \ref{lem:me:infty0}, $p$ switches to Phase 0 infinitely often. By (*), we know that $p$ eventually stops executing Action $\rsix$. So, from the code of Algorithm \ref{algo:ME}, we can deduce that $p$ eventually successively executes Actions $\rzero$, $\rone$, $\rtwo$, $\rthree$, and $\rfour$ infinitely often. Consider the first time $p$ successively executes $\rzero$, $\rone$, $\rtwo$, $\rthree$, and $\rfour$ and study the two following cases:
\BEGLIST
\item[-] {\em $p = \led$.} Then, $\val_p = 0$ and $\IDL .\Min_p = ID_p$ when $p$ executes $\rthree$ because $p$ executes a complete $\IDL$-computation between $\rzero$ and $\rone$ and $\IDL$ is snap-stabilizing for Specification \ref{spec:IDL} (page \pageref{spec:IDL}). Hence, $p$ updates $\val_p$ to 1 when executing $\rthree$ --- a contradiction.
\item[-] {\em $p \neq \led$.} Then, $\IDL .\Min_p = ID_p$ when $p$ executes $\rthree$ because $p$ executes a complete $\IDL$-computation between $\rzero$ and $\rone$ and $\IDL$ is snap-stabilizing for Specification \ref{spec:IDL} (page \pageref{spec:IDL}). Also, $p$ receives $\yes$ from $\led$ because $p$ executes a complete PIF of the message $\ask$ between $\rone$ and $\rtwo$ and $\pif$ is snap-stabilizing for Specification \ref{spec:pif} (page \pageref{spec:pif}). Hence, $p$ satisfies the predicate $Winner(p)$ when executing $\rthree$ and, as a consequence, starts a PIF of the message $\exitcs$ in Action $\rthree$. This PIF terminates when $p$ executes $\rfour$: from this point on, we have the guarantee that $\led$ has executed Action $\rseven$. Now, by $\rseven$, $\led$ increments $\val_{\led}$ --- a contradition.
\ENDLIST
\end{proof}

\begin{lemma}[Start]\label{lem:me:start}
Starting from any configuration, any process that requests the $\cs$, enters in the $\cs$ in finite time.
\end{lemma}
\begin{proof}
Assume, for the purpose of contradiction, that from a configuration $\gamma$, a process $p$ requests but never enters in the $\cs$. Then, Lemma \ref{lem:me:infty0} implies that $p$ eventually executes $\rzero$ and after executing $\rzero$, $\req_p = \rin$ holds forever ($\req_p$ is switched to $\rdone$ only after $p$ releases the $\cs$). From the code of Algorithm \ref{algo:ME}, we can then deduce that there is two possibilities after $p$ executes $\rzero$:
\BEGLIST
\item[-] $p$ no more executes $\rthree$, or
\item[-] $p$ satisfies $\neg Winner(p)$ each time it executes $\rthree$.
\ENDLIST 
Consider then the two following cases:
\BEGLIST
\item[-] {\em $p = \led$.} Then, $\val_p = 0$ eventually holds forever --- a contradiction to Lemma \ref{lem:me:incr}.
\item[-] {\em $p \neq \led$.} In this case, $p$ no more starts any PIF of the message $\exitcs$. Now, every PIF-computation terminates in finite time (termination property of Specification \ref{spec:pif}, page \pageref{spec:pif}). Hence, the ``$\BRD \langle \exitcs \rangle$ {\bf from} $p$'' event eventually no more occurs at $\led$. As a consequence, $\val_{\led}$ eventually no more switches from value $p$ to $(p + 1) \bmod (n+1)$ --- a contradiction to Lemma \ref{lem:me:incr}.
\ENDLIST
\end{proof}

\noindent By Corollary \ref{coro:me:correct} and Lemma \ref{lem:me:start}, starting from any configuration, any execution of $\me$ always satisfies Specification \ref{spec:ME}. Hence, follows:

\begin{theorem}
Protocol $\me$ is snap-stabilizing from Specification \ref{spec:ME}.
\end{theorem}

%% file: ccl.tex
\section{Conclusion}\label{sect:ccl}

We addressed the problem of {\em snap-stabilization} in message-passing systems and presented matching negative and positive results.
On the negative side, we show that {\em snap-stabilization} is impossible for a wide class of specifications --- namely, the {\em safety-distributed} specifications --- in message-passing systems where the channel capacity is finite yet unbounded. 
On the positive side, we show that {\em snap-stabilization} is possible (even for {\em safety-distributed} specifications) in message-passing systems if we assume a bound on the channel capacity. The proof is constructive, as we presented the first three snap-stabilizing protocols for message-passing systems with a bounded channel capacity. These protocols respectively solve the PIF, IDs-Learning, and mutual exclusion problem in a fully-connected network. 

On the theoretical side, it is worth investigating if the results presented in this paper could be extended to more general networks, \emph{e.g.} with general topologies, and/or where nodes are subject to permanent  \emph{aka} crash failures. On the practical side, our result implies the possibility of implementing snap-stabilizing protocols on real networks, and actually implementing them is a future challenge.

%% file: RR-6446.bbl
\begin{thebibliography}{10}

\bibitem{AB97}
Y~Afek and A~Bremler.
\newblock Self-stabilizing unidirectional network algorithms by power supply.
\newblock {\em Chicago Journal of Theoretical Computer Science}, 1998:Article
  3, 1998.

\bibitem{AB93}
Y~Afek and GM~Brown.
\newblock Self-stabilization over unreliable communication media.
\newblock {\em Distributed Computing}, 7):27--34, 1993.

\bibitem{AlpernS85}
Bowen Alpern and Fred~B. Schneider.
\newblock Defining liveness.
\newblock {\em Inf. Process. Lett.}, 21(4):181--185, 1985.

\bibitem{AlpernS87}
Bowen Alpern and Fred~B. Schneider.
\newblock Recognizing safety and liveness.
\newblock {\em Distributed Computing}, 2(3):117--126, 1987.

\bibitem{DBLP:journals/dc/AroraN05}
Anish Arora and Mikhail Nesterenko.
\newblock Unifying stabilization and termination in message-passing systems.
\newblock {\em Distributed Computing}, 17(3):279--290, 2005.

\bibitem{APV91}
B~Awerbuch, B~Patt-Shamir, and G~Varghese.
\newblock Self-stabilization by local checking and correction.
\newblock In {\em FOCS91 Proceedings of the 31st Annual IEEE Symposium on
  Foundations of Computer Science}, pages 268--277, 1991.

\bibitem{AKMPV07}
Baruch Awerbuch, Shay Kutten, Yishay Mansour, Boaz Patt-Shamir, and George
  Varghese.
\newblock A time-optimal self-stabilizing synchronizer using a phase clock.
\newblock {\em IEEE Trans. Dependable Sec. Comput.}, 4(3):180--190, 2007.

\bibitem{DBLP:conf/sss/BeinDV05}
Doina Bein, Ajoy~Kumar Datta, and Vincent Villain.
\newblock Snap-stabilizing optimal binary search tree.
\newblock In Ted Herman and S{\'e}bastien Tixeuil, editors, {\em
  Self-Stabilizing Systems}, volume 3764 of {\em Lecture Notes in Computer
  Science}, pages 1--17. Springer, 2005.

\bibitem{BCV03}
L~Blin, A~Cournier, and V~Villain.
\newblock An improved snap-stabilizing {PIF} algorithm.
\newblock In {\em DSN SSS'03 Workshop: Sixth Symposium on Self-Stabilizing
  Systems (SSS'03)}, pages 199--214. LNCS 2704, 2003.

\bibitem{BDPV99c}
A~Bui, AK~Datta, F~Petit, and V~Villain.
\newblock Snap-stabilizing {PIF} algorithm in tree networks without sense of
  direction.
\newblock In {\em SIROCCO'99, The 6th International Colloquium On Structural
  Information and Communication Complexity Proceedings}, pages 32--46. Carleton
  University Press, 1999.

\bibitem{BDPV99b}
A~Bui, AK~Datta, F~Petit, and V~Villain.
\newblock State-optimal snap-stabilizing {PIF} in tree networks.
\newblock In {\em Proceedings of the Fourth Workshop on Self-Stabilizing
  Systems}, pages 78--85, Austin, Texas, USA, June 1999. IEEE Computer Society
  Press.

\bibitem{DBLP:journals/dc/BuiDPV07}
Alain Bui, Ajoy~Kumar Datta, Franck Petit, and Vincent Villain.
\newblock Snap-stabilization and pif in tree networks.
\newblock {\em Distributed Computing}, 20(1):3--19, 2007.

\bibitem{C82}
EJH Chang.
\newblock Echo algorithms: depth parallel operations on general graphs.
\newblock {\em IEEE Transactions on Software Engineering}, SE-8:391--401, 1982.

\bibitem{CDPV02}
A~Cournier, AK~Datta, F~Petit, and V~Villain.
\newblock Snap-stabilizing {PIF} algorithm in arbitrary rooted networks.
\newblock In {\em 22st International Conference on Distributed Computing
  Systems (ICDCS-22)}, pages 199--206. IEEE Computer Society Press, 2002.

\bibitem{CDPV2003}
A~Cournier, AK~Datta, F~Petit, and V~Villain.
\newblock Enabling snap-stabilization.
\newblock In {\em 23th International Conference on Distributed Computing
  Systems (ICDCS 2003)}, pages 12--19, Providence, Rhode Island USA, May 19-22
  2003. IEEE Computer Society Press.

\bibitem{CDPV2006}
A~Cournier, S~Devismes, F~Petit, and V~Villain.
\newblock {Snap-Stabilizing Depth-First Search on Arbitrary Networks}.
\newblock {\em The Computer Journal}, 49(3):268--280, 2006.

\bibitem{CDV205}
A~Cournier, S~Devismes, and V~Villain.
\newblock Snap-stabilizing detection of cutsets.
\newblock In {\em HIPC 2005, 12th Annual IEEE Conference on High Performance
  Computing}, pages 488--497. LNCS 3769, 2005.

\bibitem{CDV05}
A~Cournier, S~Devismes, and V~Villain.
\newblock A snap-stabilizing {DFS} with a lower space requirement.
\newblock In {\em Seventh International Symposium on Self-Stabilizing Systems
  (SSS'05)}, pages 33--47, Barcelona, Spain, 2005. LNCS 3764.

\bibitem{CDV06}
A~Cournier, S~Devismes, and V~Villain.
\newblock Snap-stabilizing {PIF} and useless computations.
\newblock In {\em The Twelfth International Conference on Parallel and
  Distributed Systems (ICPADS'06)}, volume~1, pages 39--46, Minneapolis, USA,
  2006. IEEE Computer Society Press P2612.

\bibitem{DBLP:journals/jhsn/CournierDPV05}
Alain Cournier, Ajoy~Kumar Datta, Franck Petit, and Vincent Villain.
\newblock Optimal snap-stabilizing pif algorithms in un-oriented trees.
\newblock {\em J. High Speed Networks}, 14(2):185--200, 2005.

\bibitem{CDV07bis}
Alain Cournier, St{\'e}phane Devismes, and Vincent Villain.
\newblock Light enabling snap-stabilization.
\newblock {\em ACM Transactions on Autonomous and Adaptive Systems (TAAS)},
  2007.
\newblock Under soumission.

\bibitem{DDT06c}
Sylvie Delaët, Bertrand Ducourthial, and Sébastien Tixeuil.
\newblock Self-stabilization with r-operators revisited.
\newblock {\em Journal of Aerospace Computing, Information, and Communication},
  2006.

\bibitem{Dij74}
EW~Dijkstra.
\newblock Self stabilizing systems in spite of distributed control.
\newblock {\em Communications of the Association of the Computing Machinery},
  17:643--644, 1974.

\bibitem{dolev97superstabilizing}
Shlomi Dolev and Ted Herman.
\newblock Superstabilizing protocols for dynamic distributed systems.
\newblock {\em Chicago Journal of Theoretical Computer Science}, 1997.

\bibitem{GGHP00}
S~Ghosh, A~Gupta, T~Herman, and SV~Pemmaraju.
\newblock Fault-containing self-stabilizing distributed protocols.
\newblock Technical Report 00-01, Department of Computer Science, University of
  Iowa, 2000.

\bibitem{DBLP:journals/tc/GoudaM91}
Mohamed~G. Gouda and Nicholas~J. Multari.
\newblock Stabilizing communication protocols.
\newblock {\em IEEE Trans. Computers}, 40(4):448--458, 1991.

\bibitem{638828}
Rodney~R. Howell, Mikhail Nesterenko, and Masaaki Mizuno.
\newblock Finite-state self-stabilizing protocols in message-passing systems.
\newblock {\em J. Parallel Distrib. Comput.}, 62(5):792--817, 2002.

\bibitem{JADT02j}
Colette Johnen, Luc Alima, Ajoy~K. Datta, and Sébastien Tixeuil.
\newblock Optimal snap-stabilizing neighborhood synchronizer in tree networks.
\newblock {\em Parallel Processing Letters}, 12(3-4):327--340, 2002.

\bibitem{KP93a}
S~Katz and KJ~Perry.
\newblock Self-stabilizing extensions for message-passing systems.
\newblock {\em Distributed Computing}, 7:17--26, 1993.

\bibitem{DBLP:journals/jpdc/PetitV07}
Franck Petit and Vincent Villain.
\newblock Optimal snap-stabilizing depth-first token circulation in tree
  networks.
\newblock {\em J. Parallel Distrib. Comput.}, 67(1):1--12, 2007.

\bibitem{Seg83}
A~Segall.
\newblock Distributed network protocols.
\newblock {\em IEEE Transactions on Information Theory}, IT-29:23--35, 1983.

\bibitem{T01}
G~Tel.
\newblock {\em Introduction to distributed algorithms}.
\newblock Cambridge University Press, Cambridge, UK, Second edition 2001.

\bibitem{DBLP:journals/siamcomp/Varghese00}
George Varghese.
\newblock Self-stabilization by counter flushing.
\newblock {\em SIAM J. Comput.}, 30(2):486--510, 2000.

\end{thebibliography}
